# Evidence of high-temperature exciton condensation in 2D atomic double layers

Zefang Wang[1], Daniel A. Rhodes[2], Kenji Watanabe[3], Takashi Taniguchi[3], James C. Hone[2], Jie Shan[1,4,5]\*, and Kin Fai Mak[1,4,5]\*

[1]School of Applied and Engineering Physics, Cornell University, Ithaca, NY, USA
[2]Department of Mechanical Engineering, Columbia University, New York, NY, USA
[3]National Institute for Materials Science, 1-1 Namiki, 305-0044 Tsukuba, Japan
[4]Laboratory of Atomic and Solid State Physics, Cornell University, Ithaca, NY, USA
[5]Kavli Institute at Cornell for Nanoscale Science, Ithaca, NY, USA
Email: jie.shan@cornell.edu; kinfai.mak@cornell.edu

**A Bose-Einstein condensate is the ground state of a dilute gas of bosons, such as atoms cooled to temperatures close to absolute zero [1]. With much smaller mass, excitons (bound electron-hole pairs) are expected to condense at significantly higher temperatures [2-7]. Here we study electrically generated interlayer excitons in MoSe$_2$/WSe$_2$ atomic double layers with density up to $10^{12}$ cm$^{-2}$. The interlayer tunneling current depends only on exciton density, indicative of correlated electron-hole pair tunneling [8]. Strong electroluminescence (EL) arises when a hole tunnels from WSe$_2$ to recombine with electron in MoSe$_2$. We observe a critical threshold dependence of the EL intensity on exciton density, accompanied by a super-Poissonian photon statistics near threshold, and a large EL enhancement peaked narrowly at equal electron-hole densities. The phenomenon persists above 100 K, which is consistent with the predicted critical condensation temperature [9-12]. Our study provides compelling evidence for interlayer exciton condensation in two-dimensional atomic double layers and opens up exciting opportunities for exploring condensate-based optoelectronics and exciton-mediated high-temperature superconductivity [13].**

Exciton condensation is a macroscopic quantum phenomenon that has attracted tremendous theoretical and experimental interests. Condensed phases of excitons that are generated by optical pumping or through the quantum Hall states under a magnetic field have been realized in coupled semiconductor quantum wells and graphene [2-5, 14-21]. The weak exciton binding in these systems, however, limits the condensation temperature to ~ 1 K. Although high-temperature exciton condensate has been observed in 1T-TiSe$_2$ [22], the system based on a three-dimensional semimetal is limited from a future device and configurability perspective.

Two-dimensional (2D) transition metal dichalcogenide (TMD) semiconductors (MX$_2$, M = Mo and W; X = S and Se) with large exciton binding energy (~ 0.5 eV) [23, 24] and flexibility in forming van der Waals heterostructures provide an exciting platform for exploring high-temperature exciton condensation and condensate-based applications [9-12]. The *maximum* condensation temperature in TMD double layers, limited by exciton ionization in the high-density regime [25], has been predicted a fraction (~ 10 %) of the exciton binding energy [9-12], i.e. comparable to room temperature! Condensation of intralayer excitons in TMDs is, however, hindered by the short exciton lifetimes [23] and formation of competing exciton complexes at high densities, such as biexcitons [23] and electron-hole (*e-h*) droplets [26, 27]. These difficulties can be overcome by separating the electrons and holes into two closely spaced layers in a double layer structure [3, 5, 25]. Long lifetimes and still substantial binding energies (> 0.1 eV) have been demonstrated for interlayer excitons [28]. They further act like oriented electric dipoles with repulsive interactions that prevent the formation of competing exciton complexes at high densities [3, 5, 25]. Nevertheless, exciton condensation remains elusive.

Here we present experimental evidence of high-temperature interlayer exciton condensation in MoSe$_2$/WSe$_2$ double layers. The device (Fig. 1a) consists of two angle-aligned TMD monolayer crystals separated by a two- to three-layer hexagonal boron nitride (h-BN) tunnel barrier. The barrier suppresses interlayer *e-h* recombination to achieve high exciton density while maintaining strong binding [9-12]. The double layer is gated on both sides with symmetric gates that are made of few-layer graphene gate electrodes and 20-30 nm h-BN gate dielectrics. Figure 1b is an optical image of a typical device. The lateral size is about a few microns. By applying equal voltages ($V_{gate}$) to the two gates and a bias voltage ($V_{bias}$) to the WSe$_2$ layer, one can tune the carrier density in each TMD layer independently in contrast to optical pumping. In addition, the electrical method can create interlayer excitons in thermal equilibrium with the lattice, favoring condensation.

Figure 1c is a contour plot of the optical reflection contrast spectra of the double layer as a function of $V_{gate}$ for $V_{bias}$ = 0 V. The two prominent features correspond to the neutral exciton resonance of monolayer MoSe$_2$ and WSe$_2$, which lose their oscillator strengths rapidly upon doping [24]. The fall-edge sharpness provides an estimate of the disorder density in each monolayer to be a few times of $10^{11}$ cm$^{-2}$ which is consistent with the highest reported quality [29, 30]. For a large range of $V_{gate}$, both neutral exciton resonances are present; the feature disappears in WSe$_2$ with hole doping at large negative $V_{gate}$'s, and in MoSe$_2$, with electron doping at large positive $V_{gate}$'s. This is consistent with a type-II band alignment of the heterostructure [28]. In contrast, for $V_{bias}$ = 5.5 V, a *p-n* region (between the two horizontal dashed lines in Fig. 1d) opens up, in which MoSe$_2$ is electron-doped (with density *n* > 0) and WSe$_2$ is hole-doped (with density *p* > 0). The bias voltage creates an interlayer electrochemical potential difference and a steady-state non-equilibrium *e-h* double layer. The problem is well described by an electrostatic model (see Methods and Supplementary Fig. 6) that the total charge density (*n* + *p*) and the charge density imbalance (*n* − *p*) are given by $V_{bias}$ and ( $2V_{gate} − V_{bias}$ ), respectively. The *e-h* pair density *N* (the smaller of *n* and *p*) is thus given by $V_{bias} − V_{gate}$ (for *n* > *p*) or $V_{gate}$ (for *n* < *p*). Very high pair densities up to $10^{12}$ cm$^{-2}$ have been achieved.

Radiative recombination from interlayer excitons is completely suppressed in devices with h-BN barrier thicker than 1 layer due to negligible *e-h* wave function overlap [31] (Supplementary Fig. 7). We employ two other probes to study the *e-h* double layer. The large $V_{bias}$ creates a tunneling current *I* between the layers at the µA level. Enhanced tunneling is expected if the electrons and holes are bound [8, 25] (i.e. form interlayer excitons). But the large bias that is needed to open a *p-n* region excludes the possibility of observing any zero-bias Josephson-like effects in the exciton condensate [3, [14]. Electroluminescence (EL) is observed near 1.6 eV (the bright spot in Fig. 1d), which matches the energy of charged exciton in MoSe$_2$ (see Supplementary Fig. 1 for EL image). The EL arises as a hole tunnels from WSe$_2$ to MoSe$_2$ and recombines radiatively with electron in MoSe$_2$ (see Methods). The maximum EL quantum efficiency is on the order of 10$^{-4}$. EL resulted from recombination of an electron tunneled from MoSe$_2$ with hole in WSe$_2$ is also observed. It is typically much weaker presumably due to the presence of lower-energy dark exciton states in WSe$_2$ [32] and is not monitored. The measured EL intensity is directly proportional to the radiative decay rate of a hole in $n$-doped MoSe$_2$, which, unlike tunneling current under large bias, is sensitive to the emergence of a condensate. Large EL enhancement has indeed been reported in a similar situation when a hole recombines radiatively with electron in a superconductor on cooling below the critical temperature [33, 34]. Unless otherwise specified, all measurements were performed at 3.5 K. The results of two devices are presented (Fig. 1, 5 and Supplementary Fig. 1, 2, 4, 5, 6 from device 1; The rest is from device 2).

Figure 2 illustrates the tunneling characteristics of the double layer. For a fixed $V_{bias}$ (i.e. constant $n + p$), the gate dependence of $I$ shows a cusp-shaped peak centered at charge balance ($n = p$) and the current on the cusp increases with $V_{bias}$ (Fig. 2a). Similarly for a fixed $V_{gate}$ (Fig. 2b), $I$ increases with increasing $V_{bias}$ and approaches a cusp at charge balance. Beyond this point, $I$ becomes a constant, the value of which increases with increasing $V_{gate}$. Note in this region ($n < p$), the $e$-$h$ pair density is also a constant, hinting that $I$ depends only on $N$. Indeed, the simulated tunneling characteristics based on $I \propto N^{2.9}$ (Fig. 2c and 2d) are in excellent agreement with experiment (Fig. 2a and 2b). Negligible temperature dependences for $I$ were observed for the entire temperature range studied in this work (3.5 K - 180 K) (Fig. 5a).

The above observations for tunneling are not consistent with the independent particle picture. In the independent particle picture, $I$ is determined by the number of states available for tunneling [35], which is proportional to $(n + p)$ (Fig. 1d). The gate dependence of $I$ for a given $V_{bias}$ would be flat inside the $p$-$n$ region instead of a cusp-shaped peak (Fig. 2a). Tunneling here is consistent with correlated $e$-$h$ pair tunneling [8] that involves the creation and annihilation of bound $e$-$h$ pairs. The tunneling current is thus only dependent on pair density [8]. The absence of temperature dependences for $I$ is also consistent with the large exciton binding energy (> 0.1 eV, equivalent to 1200 K) predicted for interlayer excitons with ~ 1 nm h-BN separation [9, 36].

Next we turn to the EL measurements. Figure 3a displays the EL spectra for different exciton densities (at charge balance). The spectra consist of a peak with a linewidth of 10-20 meV. No significant changes in both the spectral width and the peak energy with $N$ are noted. (The detailed analysis of the EL spectra is summarized in Supplementary Fig. 2 and more discussion on the linewidth in Methods.) In contrast, the spectrally integrated EL intensity shows a critical threshold dependence on $N$ with threshold at $N_{th} \approx 0.26 \times 10^{12}$ cm$^{-2}$ (Fig. 3c). In a narrow range of $N_{th}$ the EL intensity increases by two orders of magnitude. In contrast, the tunneling current changes by only about two times in the same density range.

The threshold behavior is accompanied by a change in photon statistics revealed by the intensity correlation measurement based on a Hanbury Brown-Twiss (HBT) type setup (see Methods) (Fig. 3d). The time resolution of the setup is about 40 ps, which far exceeds the EL coherence time (~ 100 fs) estimated from the spectral width. Figure 3b displays the EL intensity correlation function $g^{(2)}(\tau)$ for different $N$'s (at charge balance), where $\tau$ is the arrival time difference between pairs of EL photons. Photon bunching ($g^{(2)}(0) > 1$) is observed near threshold with a decay time of about 1 ns. Above threshold, photon bunching is absent ($g^{(2)}(\tau) = 1$) and the EL statistics is Poissonian.

The EL enhancement is very sensitive to charge imbalance in the double layer. Figure 4 shows the EL intensity as a function of $(n - p)$ at several fixed values of $(n + p)$ (= $2N$ at charge balance). The tunneling current is included as a reference. For $N \geq N_{th}$, similar to current, EL shows a cusp-shaped peak centered at charge balance, but the EL enhancement occurs in a much narrower range than $I$. For $N < N_{th}$, the EL intensity does not follow the current strictly and does not have a cusp. Near threshold, EL is particularly sensitive to charge imbalance both in terms of intensity and photon statistics (Supplementary Fig. 3). We employ the normalized 'density width' (the full-width-half-maximum of the peak, $\delta n$, divided by the total density), $\frac{\delta n}{2N}$, to quantify the sensitivity of the two processes to charge imbalance in Fig. 3e. Whereas the current 'density width' is nearly independent of $N$, the EL 'density width' decreases sharply at threshold and remains substantially smaller than the current 'density width'. Similar EL characteristics have been observed in other devices. Supplementary Fig. 4 and 5 show results from device 1 with a higher $N_{th}$.

The observed threshold behavior on exciton density, photon bunching at threshold and absence of bunching above threshold, and high sensitivity to charge imbalance for EL are not compatible with the conventional light-emitting diode action that involves tunneling and recombination of independent charge particles. In such picture, the EL intensity is proportional to current before reaching saturation and none of these observations can be explained. The EL-threshold behavior is analogous to lasing and polariton condensation, which are known non-equilibrium continuous phase transitions [6, 7, 37-39]. In these processes, the cavity photons or the polaritons condense into a single electromagnetic mode above threshold. Photon bunching arises near threshold due to critical electromagnetic fluctuations, which disappear above threshold in the condensed phase. However, in the absence of an optical cavity that provides feedback, EL in our devices is not lasing or polariton condensation.

Instead, the EL-threshold behavior is consistent with a continuous phase transition at the critical density $N_{th}$ from an exciton gas to an exciton condensate, whose wave function consists of spatially coherent *e-h* pairs [3, 10, 14, 39]. When a 'normal' hole recombines radiatively with electron in $MoSe_2$ that is a part of a condensate, the recombination rate increases with the number of bosons in the condensate [33], i.e. a superradiant process [4, 5] (see the EL rate analysis in Methods). The situation is similar to the case when a hole is injected into a superconductor and recombines radiatively with its electron Cooper pair condensate [33, 34]. The observed photon bunching with a correlation time much longer than the coherence time near threshold corresponds to critical fluctuations and slowing down at the critical point [37, 38]. The absence of photon bunching above threshold corresponds to suppressed noise in the condensed phase. The observed strong sensitivity of the EL enhancement to charge imbalance above threshold is also consistent with exciton condensation, which requires nearly perfect *e-h* Fermi surface nesting [40-43]. In particular, a non-analytic cusp at charge balance as in Fig. 4 was predicted for exciton coherence as a function of charge imbalance [40-43]. Such sensitivity to charge imbalance disappears below threshold as expected.

Finally, we estimate the transition temperature for exciton condensation. Figure 5a displays the EL intensity and tunneling current as a function of charge imbalance at varying temperatures for a fixed pair density above threshold ($0.74 \times 10^{12}$ cm$^{-2}$). The tunneling current has a negligible dependence on charge imbalance or temperature $T$, as discussed above (Fig. 2b). In contrast, both the EL intensity and EL enhancement at charge balance decrease with increasing temperature. To exclude any potential trivial temperature effects, we use the relative EL enhancement at charge balance above the baseline to estimate the transition temperature. The top panel in Fig. 5b corresponds to the result for $N = 0.74 \times 10^{12}$ cm$^{-2}$. For $T > 100$ K, the enhancement disappears (i.e. EL behavior returns to that of a normal light-emitting diode). This value is very close to the predicted degeneracy temperature (onset of the macroscopic occupation of the ground state) for interlayer excitons in TMD double layers separated by two-layer h-BN [9]. Figure 5b also shows similar results for two other exciton densities. No clear $N$-dependence of the transition temperature can be concluded for the small range of densities investigated in this study. A systematic investigation of the $N$-dependence to test different theories (linear [9, 11] or more complicated dependences [12]) is beyond the scope of this study and deserves further investigations. We also note that all samples studied in this work are a few microns in size and further improvement on the sample size and quality is required to further investigate spatial coherence and finite sample size effects on exciton condensation [15, 19, 21].

In conclusion, we have electrically created a high-density *e-h* double layer based on 2D van der Waals heterostructures under zero magnetic field. By combining the tunneling and EL measurements, we have observed threshold dependence on exciton density, sensitive dependence on charge imbalance, and critical fluctuations for EL, which are consistent with exciton condensation. These observations persist up to ~ 100 K. Our results open up exciting opportunities for exploring exciton condensates at high-energy scales [9-12] and exciton-mediated high-temperature superconductivity [13].

**Methods**

**Device fabrication**

Dual-gated WSe$_2$/h-BN/MoSe$_2$ tunnel junctions were fabricated based on the dry-transfer method developed by Wang et al. [44] Each constituent layer of the junction and the gates was exfoliated from bulk crystals onto Si substrates with a 300-nm oxide layer. High-quality WSe$_2$ and MoSe$_2$ bulk crystals were synthesized based on the flux-growth technique [30]. Thin flakes of appropriate thickness and size were identified optically. In particular, the dark-field imaging mode was used to enhance the optical contrast for few-layer h-BN. The h-BN tunnel barrier is about 2-3 layer thick (0.6-1.0 nm), confirmed by atomic force microscopy. The identified flakes were picked up layer-by-layer by a stamp made of a thin layer of polycarbonate (PC) on polydimethylsiloxane (PDMS) supported by a glass slide. The finished stack was thermally released onto a Si substrate with pre-patterned electrodes for source, drain, top and bottom gates. The PC film was finally removed by chloroform. The device schematics and the optical image of a finished device are shown in Fig. 1a and 1b.

**Reflection contrast and electroluminescence measurements**

Reflection contrast measurements were conducted to calibrate the doping density of WSe$_2$ and MoSe$_2$ monolayers. The output of a broadband super-continuum light source was focused by a high numerical aperture (NA = 0.8) objective to a spot of about 1 μm$^2$ on the device. The reflected light was collected by the same objective, dispersed by a grating spectrometer and detected by a charge-coupled-device (CCD) camera. The reflection contrast $\delta R/R$ was obtained by normalizing the difference between the reflected light intensity from the tunnel junction and the substrate ($\delta R$) to that from the substrate ($R$). Electroluminescence (EL) from the tunnel junction was detected by the same optics. The typical integration time for EL was 0.1 s.

**Hanbury Brown and Twiss setup**

A Hanbury Brown and Twiss (HB-T) setup was employed to measure the EL intensity correlation function $g^{(2)}(\tau)$. The collected EL from the tunnel junction was split by a 50:50 beam splitter and focused onto two identical single-photon detectors. The detector outputs were fed to a time-correlated single photon counter (Picoharp 300) to record the arrival time difference between a pair of EL photons. A histogram of photon counts versus arrival time difference and the intensity correlation function can then be obtained. The instrument response time is about 40 ps.

**Electrostatics of the TMD double layer structure**

The equivalent circuit of the TMD tunnel junction is a tunnel resistor connecting the bottom and top gates. The junction capacitance can be ignored (compared with the gate capacitance $C_{gate}$) under large bias and tunneling current. The energy band alignment is shown in Fig. 1. For symmetric gating as in our experiment, we express the electron density $n$ ($> 0$) in the MoSe$_2$ layer and the hole density $p$ ($> 0$) in the WSe$_2$ layer as

$$ne = C_{gate}(V_{gate} - V_{Mo}^{off}),$$
$$pe = C_{gate}(V_{bias} - V_{W}^{off} - V_{gate}). \qquad (1)$$

Here $V_{gate}$ and $V_{bias}$ are the gate and bias voltages, respectively, $e$ is the elementary charge, and $V_{Mo}^{off}$ ($V_{W}^{off}$) is the amount of potential that is required to move the in-gap Fermi level to the conduction band minimum of MoSe$_2$ (the valence band maximum of WSe$_2$). The gate capacitance $C_{gate} = (0.89 - 1.33) \times 10^{-7}$ F/cm$^2$ is determined by the thickness of the h-BN dielectric (20-30 nm) and its dielectric constant (~ 3) [29, 30]. We derive from Eqn. (1) the *e-h* density imbalance ($n - p$), the total density ($n + p$), and the exciton density $N$ as

$$(n-p)e = C_{gate}(2V_{gate} - V_{bias} - V_{Mo}^{off} + V_{W}^{off}),$$
$$(n+p)e = C_{gate}(V_{bias} - V_{Mo}^{off} - V_{W}^{off}),$$

$$Ne = \begin{cases} C_{gate}(V_{bias} - V_{gate} - V_W^{off}) \, , & n > p \\ C_{gate}(V_{gate} - V_{Mo}^{off}) & , \, n < p \end{cases}. \tag{2}$$

The density imbalance is controlled by $(2V_{gate} - V_{bias})$, and the total density, by $V_{bias}$. We compare the electrostatic model with experiment in Supplementary Fig. 6 for three special cases: $n = p$, $n = 0$, and $p = 0$:

$$V_{gate} = \tfrac{1}{2}(V_{bias} + V_{Mo}^{off} - V_W^{off}) \quad \text{for} \quad n = p, \tag{3}$$

$$V_{gate} = V_{Mo}^{off} \quad \text{for} \quad n = 0, \tag{4}$$

$$V_{gate} = V_{bias} - V_W^{off} \quad \text{for} \quad p = 0. \tag{5}$$

The gate voltage for $n = p$ can be unambiguously determined in experiment from the cusp of the tunneling current. The gate voltages for $n = 0$ and $p = 0$ can be determined from the reflection contrast of MoSe$_2$ and WSe$_2$. We assume that when the electrochemical potential touches the band edge, the reflection contrast of the corresponding neutral exciton resonance drops to 90% of its maximum value. We obtain $V_{Mo}^{off} \approx 2.3$ V and $V_W^{off} \approx 2.0$ V from the best fit in Supplementary Fig. 6b.

**Electroluminescence (EL) rate equation and spectral width**

The rate equation for the minority hole density $p_0$ in the $n$-doped MoSe$_2$ layer can be expressed as

$$\frac{dp_0}{dt} = \frac{I}{2Ae} - p_0 \gamma_{tot}. \tag{6}$$

The first term is the hole pumping rate with $I$ denoting the tunneling current, $A$ the tunnel junction area, and $e$ the elementary charge. Since tunneling here is a correlated pair-wise process (Fig. 2), the hole pumping rate is determined by half of the tunneling current $I/2$. The second term is the total hole decay rate in MoSe$_2$ with $\gamma_{tot} = \gamma_r + \gamma_{nr} + \gamma_C$, where $\gamma_r$, $\gamma_{nr}$ and $\gamma_C$ represent contributions from the radiative recombination, non-radiative recombination in MoSe$_2$, and tunneling out to the electrode, respectively. In the steady state, $\frac{dp_0}{dt} = 0$, equation (6) yields $p_0 = \frac{I}{2Ae\gamma_{tot}}$. The integrated EL power normalized by tunneling current is thus determined by the ratio of the radiative rate to the total decay rate

$$p_0 \gamma_r A E_X / I = \frac{1}{2e} \frac{\gamma_r}{\gamma_{tot}} E_X. \tag{7}$$

Here $E_X$ is the EL photon energy (~ 1.6 eV). In the vicinity of exciton condensation, the large enhancement observed in the normalized EL power (Fig. 3c) is dominated by the enhancement in $\gamma_r$ since the EL quantum yield is typically very small (i.e. $\frac{\gamma_r}{\gamma_{tot}} \ll 1$). The radiative recombination rate $\gamma_r$ is determined by the transition dipole matrix element between the initial and final state, which are many-body quantum states in the exciton condensate phase [33]. The radiative rate is enhanced by the number of spatially coherent excitons in the condensate, similar to the phenomenon of super-radiance.

Unlike in lasing and polariton condensation, significant EL spectral line narrowing does not necessarily accompany the threshold intensity dependence on exciton density. In lasing or polariton condensation, while the carrier decay rate, thermal broadening and inhomogeneous broadening contribute to the spectral width below threshold, the spectral width above threshold is determined only by the photon decay rate in the cavity, which is typically much smaller, giving rise to significant spectral line narrowing. In the absence of an optical cavity, the EL spectral width below and above threshold in our devices is mainly determined by $\gamma_{tot}$, which is not necessarily significantly suppressed above threshold. There is therefore enhanced *spatial* coherence but not necessarily much enhanced *temporal* coherence in the condensed phase. The situation is similar to the case of a hole recombining radiatively with electron in a superconductor [33, 34]. Significantly enhanced EL intensity due to the formation of a Cooper pair condensate was observed without significant line narrowing below the critical temperature.

# Figures

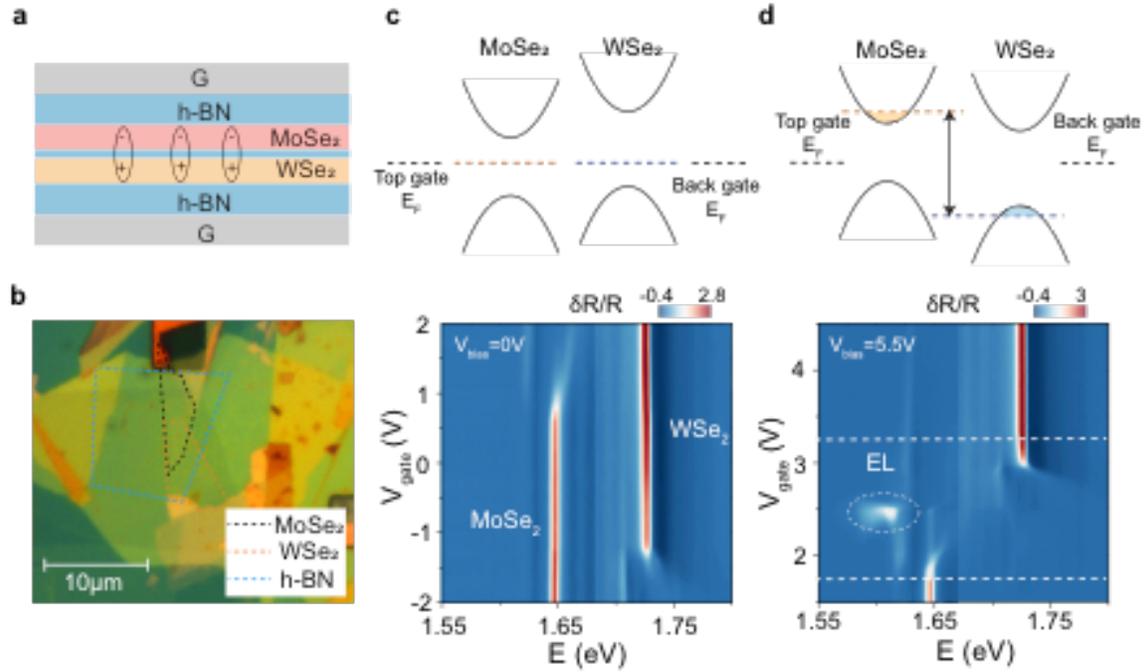

**Figure 1 | Electrical generation of high-density exciton gases. a, b,** Schematics (**a**) and optical micrograph (**b**) of a MoSe$_2$/WSe$_2$ double layer with an h-BN barrier and symmetric top and bottom gates made of h-BN and graphite (G). Interlayer excitons (electrons in MoSe$_2$ and holes in WSe$_2$) are created by electrical biasing and gating. **c,** Contour plot of the reflection contrast as a function of photon energy E and gate voltage $V_{gate}$ (identical on both gates) under zero bias voltage (lower panel). It is consistent with a type-II band alignment of the double layer (upper panel). The Fermi levels (dashed lines) are aligned for the two TMD layers. The MoSe$_2$ and WSe$_2$ neutral exciton resonances coexist in a large range of $V_{gate}$ corresponding to the Fermi level inside the gap of both layers. **d,** Same as **c** for $V_{bias}$ = 5.5 V. A gap is opened for the MoSe$_2$ and WSe$_2$ neutral exciton resonances (between the white dashed lines, corresponding to 90% of the peak reflection contrast) (lower panel). It is consistent with the band alignment (upper panel). The bias voltage splits the MoSe$_2$ and WSe$_2$ electrochemical potentials (orange and blue dashed lines). The circled bright spot in the lower panel is EL from the tunnel junction. (Data from Device 1)

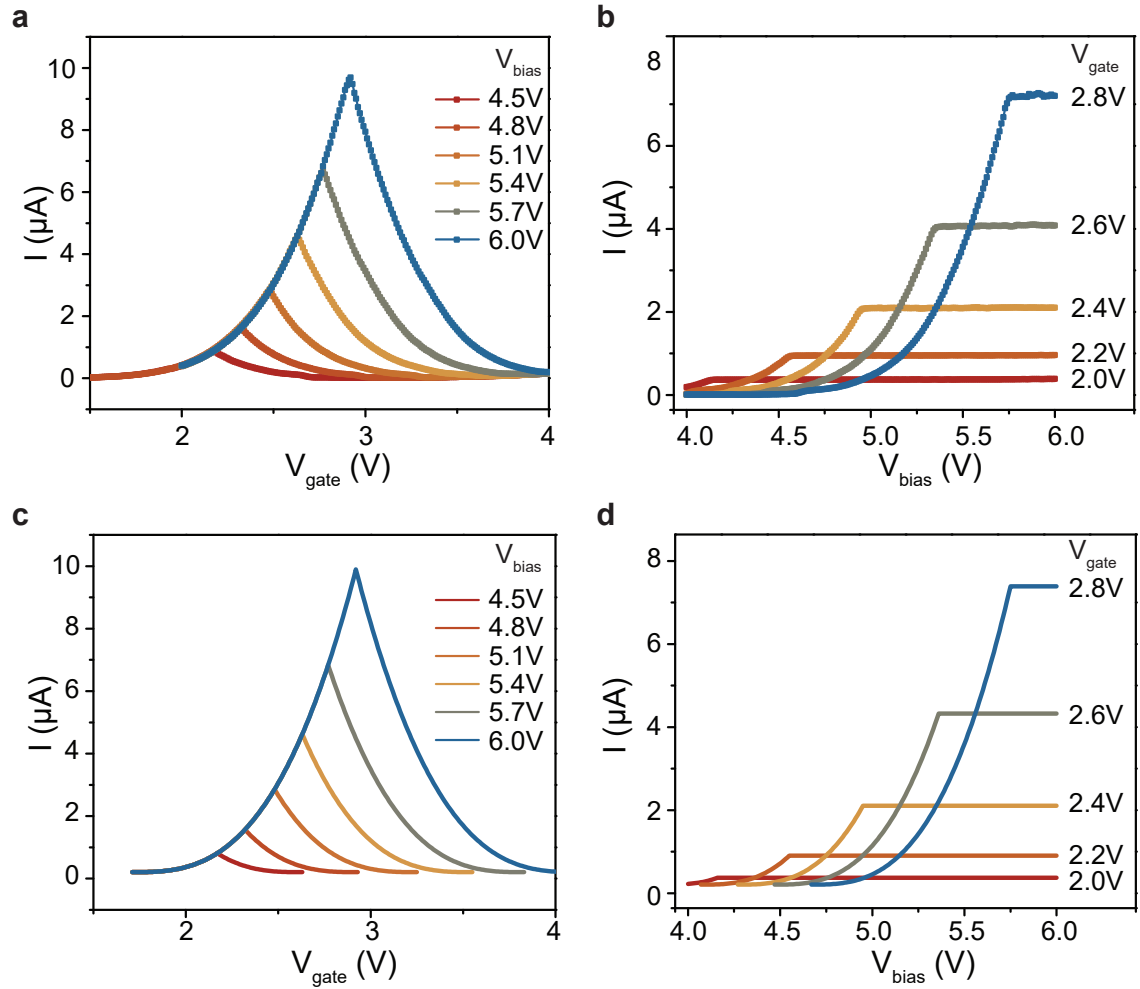

**Figure 2 | Tunneling characteristics and correlated pair tunneling. a, b,** Tunneling current $I$ of the double layer as a function of gate voltage at varying bias voltages (**a**) and as a function of bias voltage at varying gate voltages (**b**). **c, d,** The corresponding tunneling characteristics simulated based on $I = \alpha N^\beta$, where $N$ is the exciton density, $\alpha$ and $\beta$ are the free fitting parameters for the entire family of dependences. The best-fit result for exponent $\beta$ is 2.9. (Data from Device 2)

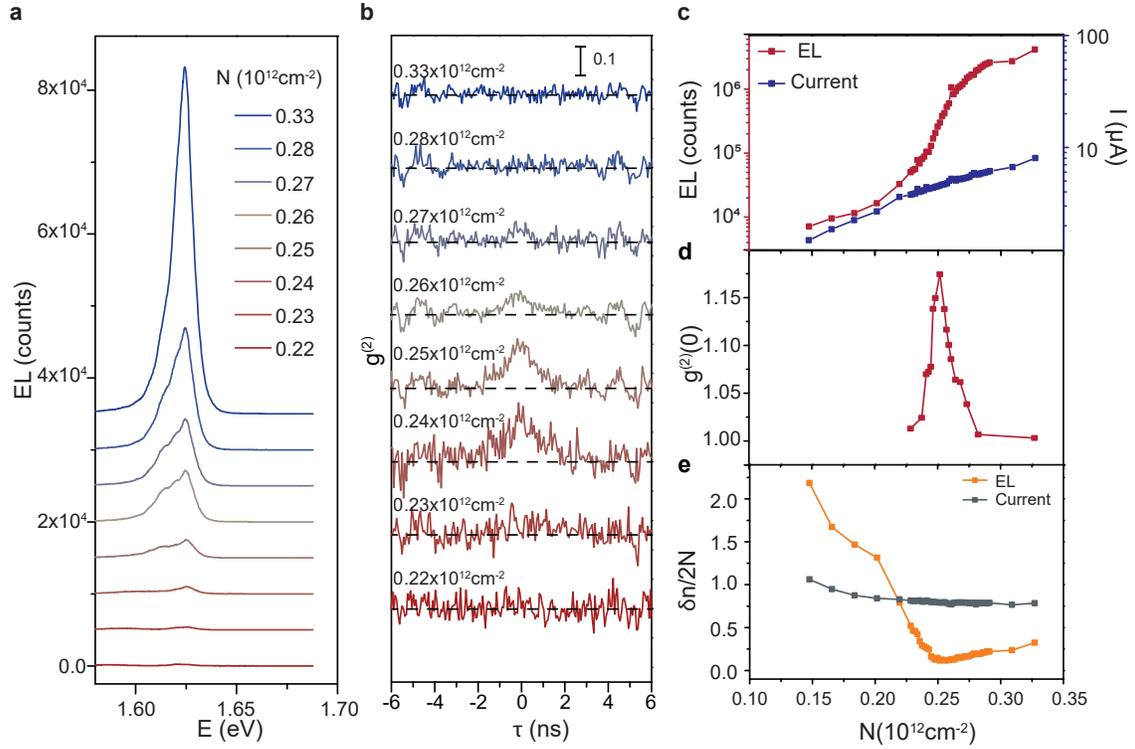

**Figure 3 | Threshold dependence of EL on exciton density at 3.5 K. a,** EL spectra at varying exciton densities (at charge balance). **b,** EL intensity correlation function at the same exciton densities as in **a**. Dashed lines note $g^{(2)} = 1$. The curves in **a**, **b** are displaced vertically for clarity. **c,-e,** Exciton density dependence of the integrated EL intensity and tunneling current (**c**), the EL intensity correlation at zero time delay (**d**), and the EL and tunneling 'density width' (**e**). The 'density widths' are extracted from data in Fig. 4 as described in the text. (Data from Device 2)

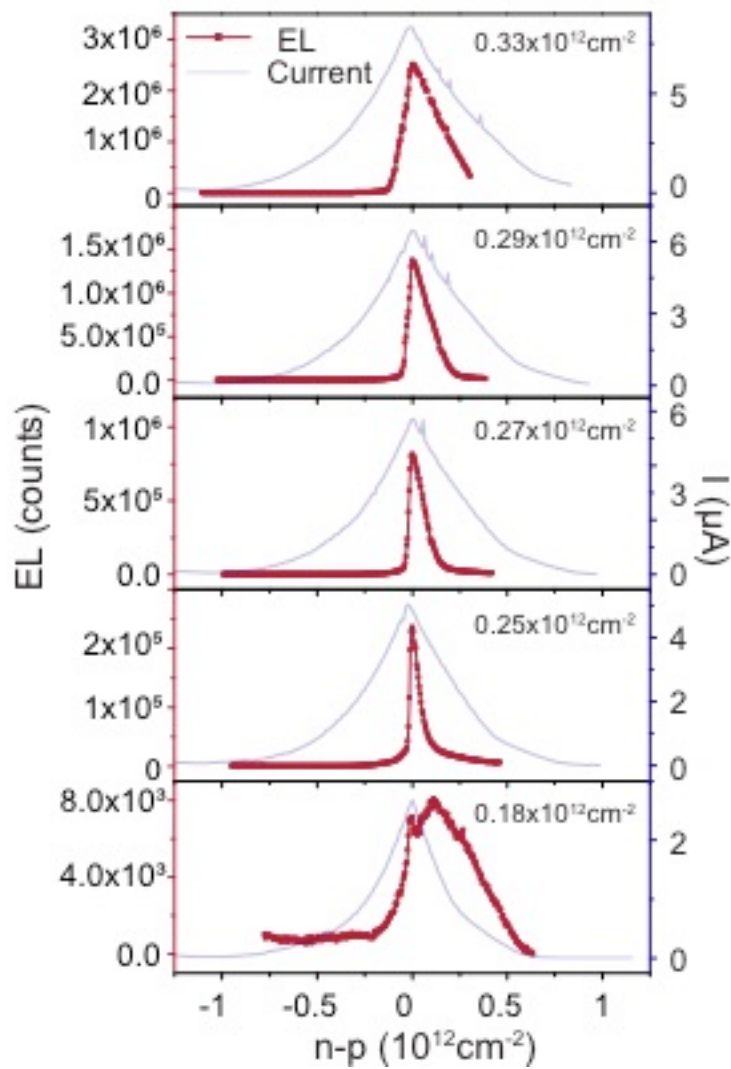

**Figure 4 | Effect of electron-hole density imbalance.** Integrated EL intensity (left axis) and tunneling current (right axis) as a function of electron-hole density imbalance ($n - p$) at varying total densities. Densities shown are the exciton density measured at charge balance. Above threshold, a cusp-shaped peak centered at charge balance is observed for both EL and tunneling. The EL 'density width' is significantly narrower than the tunneling 'density width'. An asymmetric dependence for the EL is observed, presumably due to the asymmetric disorder density in $MoSe_2$ and $WSe_2$. (Data from Device 2)

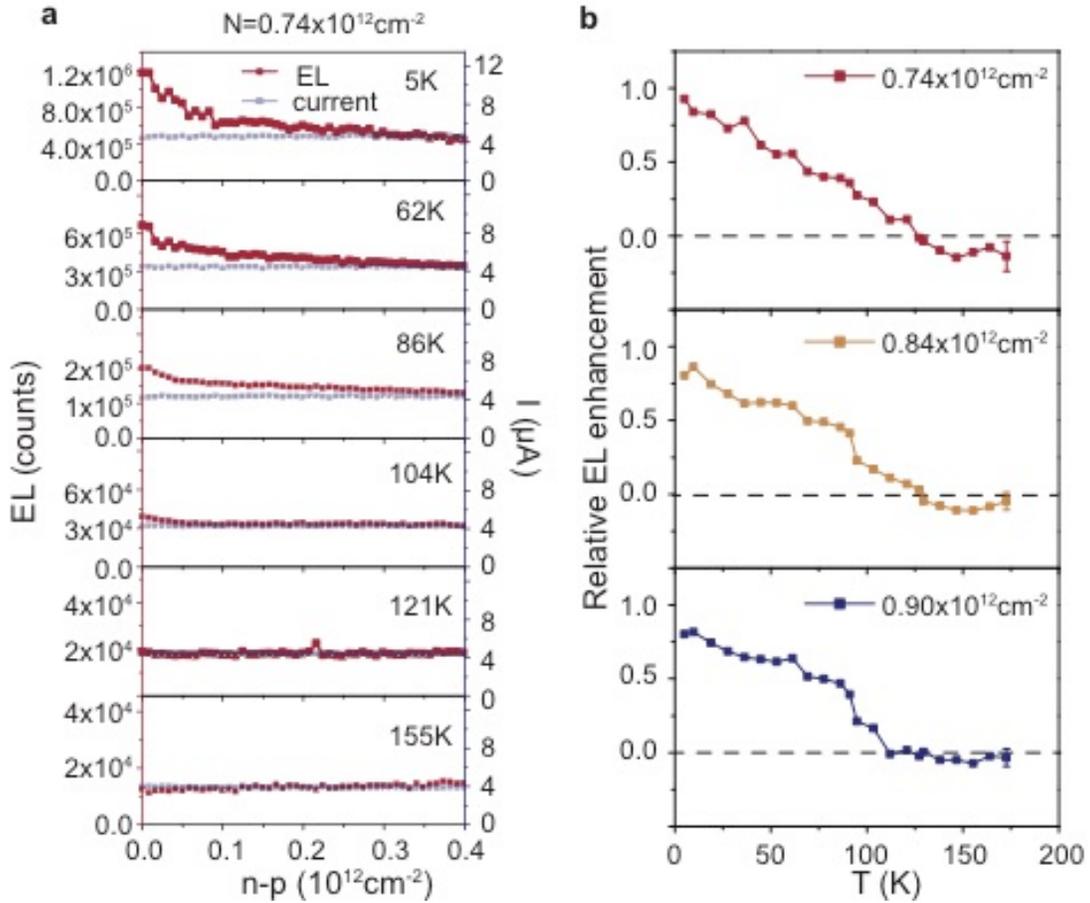

**Figure 5 | Temperature dependence of EL. a,** EL intensity (left axis) and tunneling current (right axis) as a function of charge imbalance ($n - p$) at varying temperatures for a fixed exciton density $N$ above threshold ($0.74 \times 10^{12}$ cm$^{-2}$). The EL enhancement at charge balance decreases with increasing temperature and disappears slightly above 100 K. **b,** The temperature dependence of the relative EL enhancement at charge balance above the baseline for $N = 0.74 \times 10^{12}$ cm$^{-2}$ (top), $0.84 \times 10^{12}$ cm$^{-2}$ (middle), and $0.90 \times 10^{12}$ cm$^{-2}$ (bottom). The typical errors for each density were estimated from the fluctuations of the EL intensity. The transition temperature is estimated from the value at which the enhancement disappears (dashed lines). (Data from Device 1)